\documentclass[letter]{aa} % for the letters 
\usepackage{graphicx}
\usepackage{txfonts}
\usepackage{hyperref}
\newcommand{\degg}{\hbox{$^\circ$}}

\newcommand{\xmm}{{\it XMM-Newton}}

\newcommand{\ls}
{\mathrel{\hbox{\rlap{\hbox{\lower4pt\hbox{$\sim$}}}\hbox{$<$}}}}
\newcommand{\gs}
{\mathrel{\hbox{\rlap{\hbox{\lower4pt\hbox{$\sim$}}}\hbox{$>$}}}}

\begin{document}

   \title{Variable oxygen emission from the accretion disk of Mrk\,110.}

   \author{J. N.\ Reeves \inst{1,2}  
         \and 
          D.\ Porquet \inst{3} 
          \and
          V. Braito \inst{2,1}
          \and
           N.\ Grosso\inst{3}
           \and            
           A.\ Lobban\inst{4}
          }
   \institute{Department of Physics, Institute for Astrophysics and Computational
Sciences, The Catholic University of America, Washington, DC 20064, USA 
              \email{james.n.reeves456@gmail.com}
        \and  INAF, Osservatorio Astronomico di Brera, Via Bianchi 46 I-23807 Merate (LC), Italy       
        \and  Aix Marseille Univ, CNRS, CNES, LAM, Marseille, France
        \and European Space Agency (ESA), European Space Astronomy Centre (ESAC), E-28691 Villanueva de la Cañada, Madrid, Spain 
              }
   \date{Received , 2020; accepted , 2020}

% \abstract{}{}{}{}{} 
% 5 {} token are mandatory
 
 \abstract{Six {\it XMM-Newton} observations of the bright narrow line Seyfert 1, Mrk 110, from 2004--2020, are presented. The analysis of the grating spectra from the Reflection Grating Spectrometer (RGS) 
 reveals a broad component of the He-like Oxygen (O\,\textsc{vii}) line, with a full width at half maximum (FWHM) of $15900\pm1800$\,km\,s$^{-1}$ measured in the 
 mean spectrum. The broad O\,\textsc{vii} line in all six observations can be modelled 
 with a face-on accretion disk profile, where from these profiles the inner radius of the line emission is inferred to lie between about 20--100 gravitational radii from the black hole. The derived inclination angle, of about 10 degrees, is  consistent with studies of the optical Broad Line Region in Mrk\,110. The line also appears variable and for the first time, a significant correlation is measured between the O\,\textsc{vii} flux and the continuum flux from both the RGS and EPIC-pn data. Thus the line responds to the continuum, being brightest when the continuum flux is highest, similar to the reported behaviour of the optical He\,\textsc{ii} line. The density of the line emitting gas is estimated to be $n_{\rm e}\sim10^{14}$\,cm$^{-3}$, consistent with an origin in the accretion disk.}
 
    \keywords{X-rays: individuals: Mrk\,110 -- Galaxies: active --
     (Galaxies:) quasars: general -- Radiation mechanism: general -- Accretion, accretion
     disks -- }
   \maketitle
%
%________________________________________________________________

\section{Introduction}

Studying the X-ray spectral components in active galactic nuclei (AGN), from the soft excess \citep{Czerny87, Gierlinski04} to the Fe K line \citep{Nandra07, Patrick12, Risaliti13}, can infer the properties of the innermost accretion disk and its immediate environment. 
In the so-called bare AGN, our line of sight intercepts little or no 
X-ray absorption, for example for \object{Fairall\,9} \citep{Emman11, Lohfink16} or \object{Ark\,120}, \citep{Vaughan04, Reeves16, Porquet18}. 
They provide the clearest view of the disk-corona system, where most of the
accretion power is radiated. As a result, their X-ray emission line profiles can 
be measured with less ambiguity, avoiding the complexities introduced from modelling an intervening warm absorber \citep{Miller09}.  
 
The X-ray bright, bare AGN, \object{Mrk 110}, has been classified as a narrow line Seyfert 1 (NLS1) due to its relatively narrow broad optical permitted lines 
(H\,$\beta$ FWHM $<$2000 km\,s$^{-1}$; \citealt{Bischoff99,Grupe04}). Its has no intrinsic X-ray absorption and displays a prominent soft X-ray excess below 2\,keV, 
with a strong, broadened O\,\textsc{vii} emission line \citep{Boller07}. 
However, Mrk\,110 has an unusually large [O\,III]$\lambda$5007/H$\beta$ ratio and very weak Fe\,II emission, which is more
consistent with broad line Seyfert 1s (BLS1s) rather than NLS1s \citep{Veron-cetty07} and indicates a low Eddington ratio \citep{Boroson92, Grupe04, Veron-cetty07}.
Another peculiar characteristic of Mrk\,110 is that it has a rather large black hole mass, as deduced from the detection of gravitational redshifted emission in the variable components of the broad optical lines, with a mean value of M$_{\rm grav}$= 14$\pm$3$\times$10$^{7}$\,M$_{\odot}$ \citep{Kollatschny03}. 
Comparing this to the virial mass estimate derived from reverberation mapping (M$_{\rm vir}$= 1.8$\pm$0.4$\times$10$^{7}$\,M$_{\odot}$),    
\cite{Kollatschny03} deduced a face-on orientation ($\theta=21\pm5\degg$) and a disk-like broad line region (BLR) geometry. 
\cite{Liu17} have recently confirmed this nearly pole-on view of the disk-like BLR in Mrk\,110, with an inclination of $\theta=7-10\degg$. 
The face-on orientation of Mrk\,110 may explain its unusual optical characteristics and its bare X-ray spectrum. 

Here, we report on six soft X-ray spectra of Mrk\,110, obtained from 2004--2020 with the Reflection Grating Spectrometer 
(RGS: \citealt{denHerder01}) on-board \xmm. A broad O\,\textsc{vii} emission line is observed, confirming its original detection in the first \xmm\ observation in 2004 by \citet{Boller07}. 
From modelling the line profile, we show that it may originate from a face-on accretion disk, on scales of tens of gravitational radii from the black hole.  Furthermore, for the first time we demonstrate that the O\,\textsc{vii} line is variable and correlates with the continuum flux.  

\section{Observations and data analysis}\label{sec:obs}

\begin{figure}
\begin{center}
\hspace{-0.5cm}
\rotatebox{0}{\includegraphics[width=9cm]{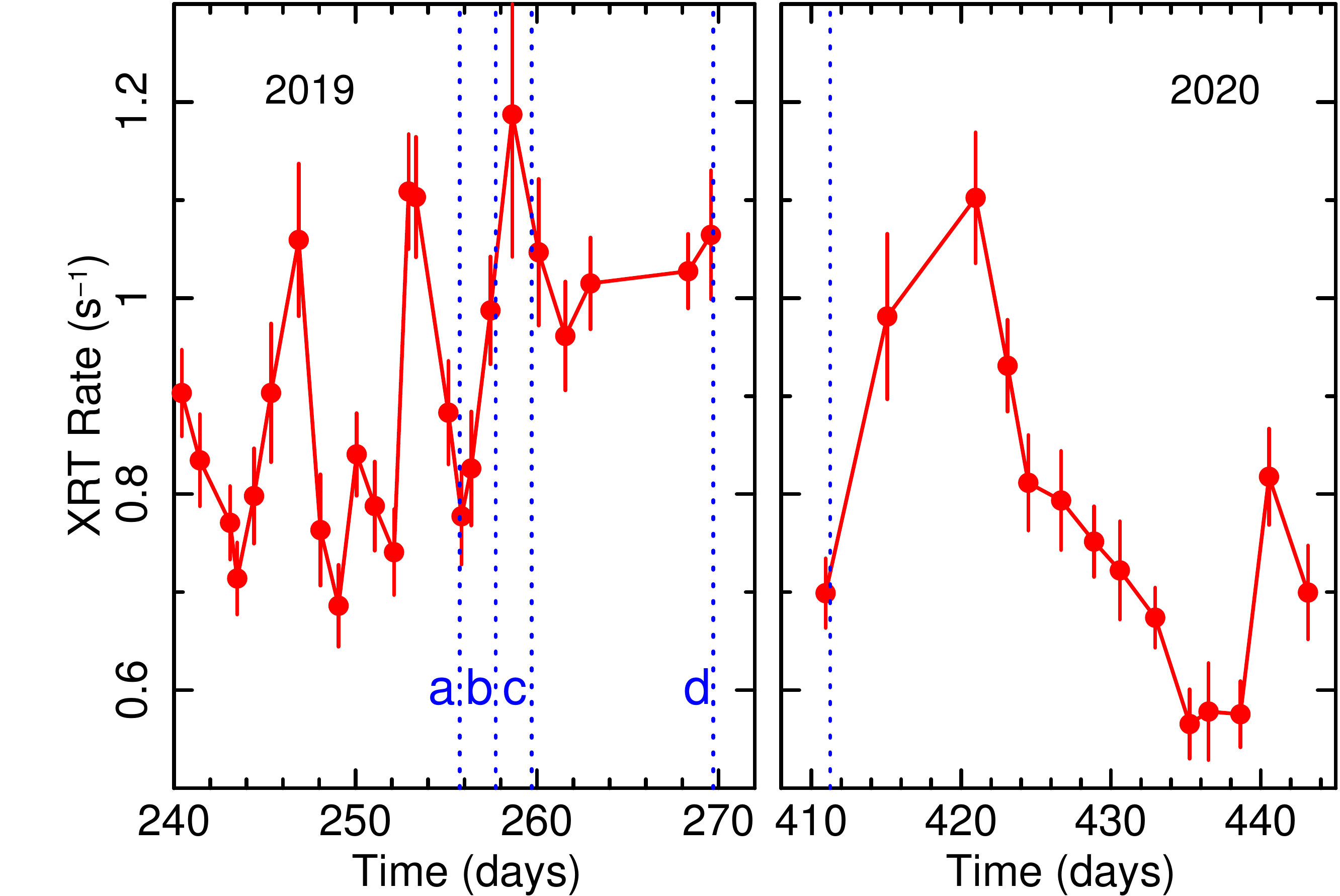}}
\end{center}
\caption{{\it Swift} XRT (0.3--2.0 keV) light curve of Mrk\,110, showing the portions closest to the times of five of the \xmm\ observations.  The \xmm\ start times are marked with vertical dotted lines, 
with four occurring in November 2019 ($a-d$ respectively) and one in April 2020 at the start of the 2020 {\it Swift} monitoring period. Mrk\,110 shows factor of two soft X-ray variability. 
Time is plotted from the beginning of the 2019--20 {\it Swift} monitoring, which commenced on February 20, 2019.}
\label{fig:swiftzoom}
\end{figure}

Mrk\,110 was observed six times with \xmm, see Table~\ref{tab:log}. 
Four observations were performed in November 2019 (hereafter 2019a--d), with one six months later in April 2020. Two of the observations (2019d and 2020, Table~\ref{tab:log}) were also performed simultaneously with {\it NuSTAR} (Principle Investigator (PI) D. Porquet)
and an analysis of their broad band X-ray spectra will be presented in a subsequent work.    
The new observations coincided with a 2019--2020 {\it Swift} monitoring campaign on Mrk\,110 (PI A. Lobban). Figure~\ref{fig:swiftzoom} shows the portions of the {\it Swift} XRT 0.3--2.0 keV band light curve which coincided with the five \xmm\ observations in 2019--2020. 
The 2019a and 2020 observations coincided with lower fluxes, while all three 2019a--c observations (PI E. Costantini) were each separated by only 2 days, encompassing a mini X-ray flare.  
The 2019d observation was 10 days later and on a flat part of the curve. The 2004 exposure was the brightest one and occurred prior to any {\it Swift} monitoring. 
The RGS light curve is shown in Appendix~A. 

\begin{table}[t!]
\caption{Log of the {\sl XMM-Newton} RGS observations of Mrk~110.}
%\label{to}
\begin{tabular}{llcccc}
\hline \hline
Sequence & Obs Date & Exp$^{a}$ & Rate$^{a}$ & Flux$^{b}$\\
\hline
0201130501 & 2004/11/15 & 47.1 & $0.803\pm0.003$ & 3.33 \\
0840220701 & 2019/11/03 & 42.1 & $0.398\pm0.002$ & 1.77 \\
0840220801 & 2019/11/05 & 38.4 & $0.584\pm0.003$ & 2.58 \\
0840220901 & 2019/11/07 & 39.3 & $0.619\pm0.003$ & 2.73 \\
0852590101 & 2019/11/17 & 39.2 & $0.629\pm0.003$ & 2.79 \\
0852590201 & 2020/04/06 & 47.2 & $0.472\pm0.002$ & 2.11 \\
\hline
\hline
\end{tabular}
\label{tab:log}
\flushleft
\small{\textit{Notes.} $^a$Net exposure (in ks) and count rate, per RGS detector. $^b$Total flux over the 0.3-2.0\,keV RGS band 
in units of $\times10^{-11}$\,erg\,cm$^{-2}$\,s$^{-1}$.} 
\end{table}

To study further the O\,\textsc{vii} line and its variability, first-order RGS spectra from all six observations of Mrk\,110 were reduced using the \textsc{rgsproc} script as part of Science Analysis System (SAS) v18.0.0. RGS\,1 and RGS\,2 spectra for each observation were combined and binned to $\Delta\lambda=0.08$\,\AA, sampling the full width at half maximum (FWHM) spectral resolution of the RGS. Spectra were fitted over the 6--35\,\AA\ range. The net count rates, fluxes, and exposures (per RGS module) are reported in Table~\ref{tab:log}.

The Galactic column density towards Mrk\,110 is $N_{\rm H}$=1.30$\times$10$^{20}$\,cm$^{-2}$ \citep{Kalberla05}.  
We used the X-ray absorption model {\sc tbnew (v2.3.2)} from
\cite{Wilms00}, assuming their interstellar medium (ISM) elemental abundances and
the cross-sections from \cite{Verner96}. 
The C-statistic \citep{Cash79} was used for the RGS fits and all errors are quoted at 90 percent confidence 
for one interesting parameter ($\Delta C=2.71$). 
Parameters are given in the AGN rest-frame at $z=0.03529$. 

\section{Emission line analysis}

To model the continuum of each of the six RGS spectra, a simple broken power-law function was adopted. This allows for the photon index to become steeper at lower energies.  
This results in a soft X-ray photon index of between $\Gamma=2.45\pm0.03$ (2004) and $\Gamma=2.12\pm0.05$ (2019a), with a trend for the photon index to steepen slightly with increasing flux between observations. Above a break energy of $E_{\rm break}=0.90\pm0.08$\,keV, the photon indices are somewhat harder, ranging from $\Gamma=2.11\pm0.06$ (2004) to $\Gamma=1.95\pm0.07$ (2019a).  There is no ionised absorption apparent towards Mrk\,110; an upper limit of $N_{\rm H}<2.6\times10^{20}$\,cm$^{-2}$ was found for a warm absorber with an 
ionisation parameter between $\log(\xi/\,{\rm erg}\,{\rm cm}\,{\rm s}^{-1})=1-3$. 

\begin{figure}
\begin{center}
\hspace{-0.7cm}
\rotatebox{0}{\includegraphics[width=9.5cm]{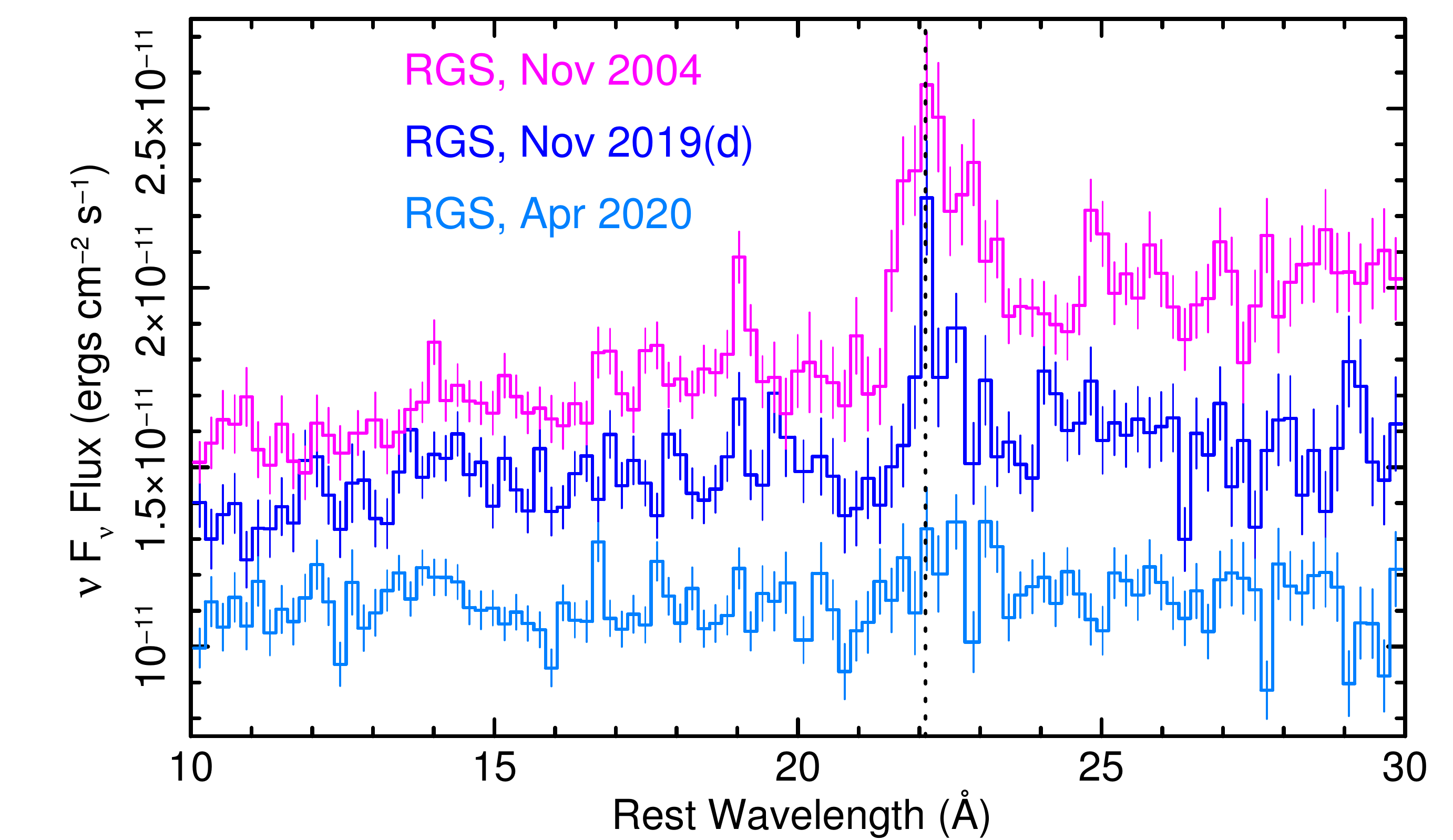}}
\end{center}
\caption{\xmm\ RGS spectra of Mrk\,110, fluxed against a $\Gamma=2$ power-law, showing the 2004, 2019d, and 2020 observations spanning the typical flux range of the AGN. The expected wavelength of the O\,\textsc{vii} forbidden line (22.1\AA) is shown by the dotted vertical line. Strong and broad O\,\textsc{vii} emission is seen in the brightest 2004 observation, which diminishes towards the 2019d and 2020 (faintest) observations. The continuum also becomes somewhat softer with increasing flux. The spectra are re-binned to $\Delta\lambda=0.2\AA$ bins for clarity.}
\label{fig:fluxed}
\end{figure} 

Clear deviations due to line emission are present in the region around the He-like O\,\textsc{vii} complex between 21--24\,\AA. 
An initial view of three of the spectra (2004, 2019d, and 2020) is shown in Figure~\ref{fig:fluxed}. The prominent O\,\textsc{vii} line near 22\,\AA\ is present in all three spectra; however, it appears strongest in the high flux 2004 observation and weakest in the low flux 2020 observation. The emission also appears broadened. A composite line profile, consisting of broad and narrow Gaussian components, was fitted to the oxygen emission. The intensity of the broad component appears strongly variable. In 2004, the broad O\,\textsc{vii} line had a flux of $N=(4.1\pm0.8)\times10^{-4}$\,photons\,cm$^{-2}$\,s$^{-1}$, which is diminished by about a factor of five to $N=(0.7\pm0.3)\times10^{-4}$\,photons\,cm$^{-2}$\,s$^{-1}$ in the low flux 2020 observation. Considering these three observations, the line is broadest in the 2004 spectrum, with a width of $\sigma=12.1^{+2.3}_{-1.9}$\,eV, corresponding to a FWHM velocity width of $v_{\rm FWHM}=15300^{+2900}_{-2400}$\,km\,s$^{-1}$. In the 2019d spectrum, its width is 
$\sigma=7.7^{+2.8}_{-2.1}$\,eV (or $v_{\rm FWHM}=9800^{+3600}_{-2700}$\,km\,s,$^{-1}$), while only in the 2020 observation where the emission is weak is the broadening not confirmed ($\sigma<4$\,eV). The energy centroid is also redshifted compared to that expected from either the forbidden, inter-combination, or resonance lines at 561\,eV, 568.5\,eV, and 574\,eV, respectively; for example, for the 2004 observation, $E=555.8\pm2.9$\,eV. This and the large velocity widths, which are significantly higher than for the optical BLR of Mrk\,110 \citep{Kollatschny03, Liu17}, suggest an origin in the accretion disk.

\begin{table}[t!]
\caption{Results from O\,\textsc{vii} emission line fitting.}
%\label{to}
\begin{tabular}{lccccc}
\hline \hline
OBS & $R_{\rm in}$$^{a}$ & $N_{\rm kyrline}$$^b$ & $N_{\rm PN}$$^c$ & $N_{\rm 0.5\,keV}^{d}$ & $\Delta C^{e}$\\
\hline
2004 & $31^{+5}_{-4}$ & $4.1\pm0.7$ & $4.3\pm0.6$ & $4.89\pm0.10$ & 98.1\\
2019a & $106^{+81}_{-61}$ & $1.5\pm0.5$ & $<1.2$ & $2.34\pm0.05$ & 28.4\\
2019b & $17^{+7}_{-6}$ & $1.8\pm0.6$ & $1.6\pm0.7$ & $3.75\pm0.07$ & 18.2\\
2019c & $37^{+7}_{-9}$ & $2.1\pm0.6$ & $1.6\pm0.8$ & $3.82\pm0.08$ & 26.9\\
2019d & $85^{+22}_{-13}$ & $2.4\pm0.6$ & $2.5\pm0.5$ & $3.92\pm0.08$ & 41.1\\
2020 & $83^{+32}_{-63}$ & $1.2\pm0.5$ & $1.2\pm0.5$ & $2.83\pm0.06$ & 15.5\\
\hline
\hline
\end{tabular}
\label{tab:oxygen}
\flushleft
\small{\textit{Notes.} $^a$Inner disk radius in units of $R_{\rm g}$. 
$^{b}$Line flux, fitted with the \textsc{kyrline} model to the RGS spectra. Units are $\times10^{-4}$\,photons\,cm$^{-2}$\,s$^{-1}$. 
$^c$Line flux measured from the EPIC-pn spectra, with a Gaussian model. Units are as stated previously. 
$^{d}$Monochromatic continuum flux measured at 0.5\,keV in units $\times10^{-2}$\,photons\,cm$^{-2}$\,s$^{-1}$\,keV$^{-1}$. 
$^e$Improvement in C-statistic upon adding the \textsc{kyrline} component.} 
\end{table}

In contrast, a narrow ($\sigma<1$\,eV) component occurs at an energy of $E=561.1\pm0.5$\,eV (or 22.1\,\AA) and is likely associated with the O\,\textsc{vii} forbidden line from distant, low density, photoionised gas \citep{Kinkhabwala02}. Its flux, of $N=(2.9\pm1.2)\times10^{-5}$\,photons\,cm$^{-2}$\,s$^{-1}$, is consistent with being constant across all observations, which suggests an origin distant from the black hole.   
The intensity is much lower than the broad component. 
Similarly, a weak narrow resonance line is also found at $E=573.1\pm0.7$\,eV and with a similar intensity. 
The only other narrow component is from the H-like line of O\,\textsc{viii} Ly$\alpha$, which occurs at $E=653.7\pm0.6$\,eV (or $\sim19$\,\AA). 
These weak narrow line components, as well as the strong broad O\,\textsc{vii} profile, are most evident in the mean profile, as is shown in Figure~\ref{fig:profiles} (upper panel). The mean broad line has a redshifted centroid energy of $E=551.1\pm1.6$\,eV, a FWHM width of 
$15900\pm1800$\,km\,s$^{-1}$, and a flux of $N=(1.7\pm0.2)\times10^{-4}$\,photons\,cm$^{-2}$\,s$^{-1}$.   
   
\subsection{Accretion disk line profiles}

The \textsc{xspec} \textsc{kyrline} model was then applied \citep{Dovciak04} to fit the O\,\textsc{vii} emission with profiles computed from the surface of an axi-symmetric accretion disk, including the special and general relativistic effects expected from close to the black hole. Identical results were also found using the \textsc{relline} model of \citet{Dauser10}. The line rest energy was restricted to lie between 561--574\,eV and the best fit value was found to be $565\pm4$\,eV, intermediate between the forbidden and intercombination lines. It is important to note that the latter may dominate in the denser environment of the disk \citep{Porquet00}. A disk annulus was assumed, where the outer radius is $R_{\rm out}= R_{\rm in} + 50R_{\rm g}$ (and $R_{\rm g}=GM/c^2$ is the gravitational radius).  
However, the modelling is not sensitive to the outer disk radius and a value of $R_{\rm out}=400R_{\rm g}$ resulted in only a marginally worse fit (by $\Delta C=4.2$).
An emissivity law of $R^{-3}$ was assumed. The profiles obtained were not dependent on the black hole specific angular momentum $a$ as the emission was found to typically arise from tens of gravitational radii. Thus $a=0$ was assumed. The constant narrow components were included as above; however, as a result of their low flux, they have a negligible impact on the derived accretion disk line parameters.

\begin{figure}
\begin{center}
\rotatebox{-90}{\includegraphics[height=9cm]{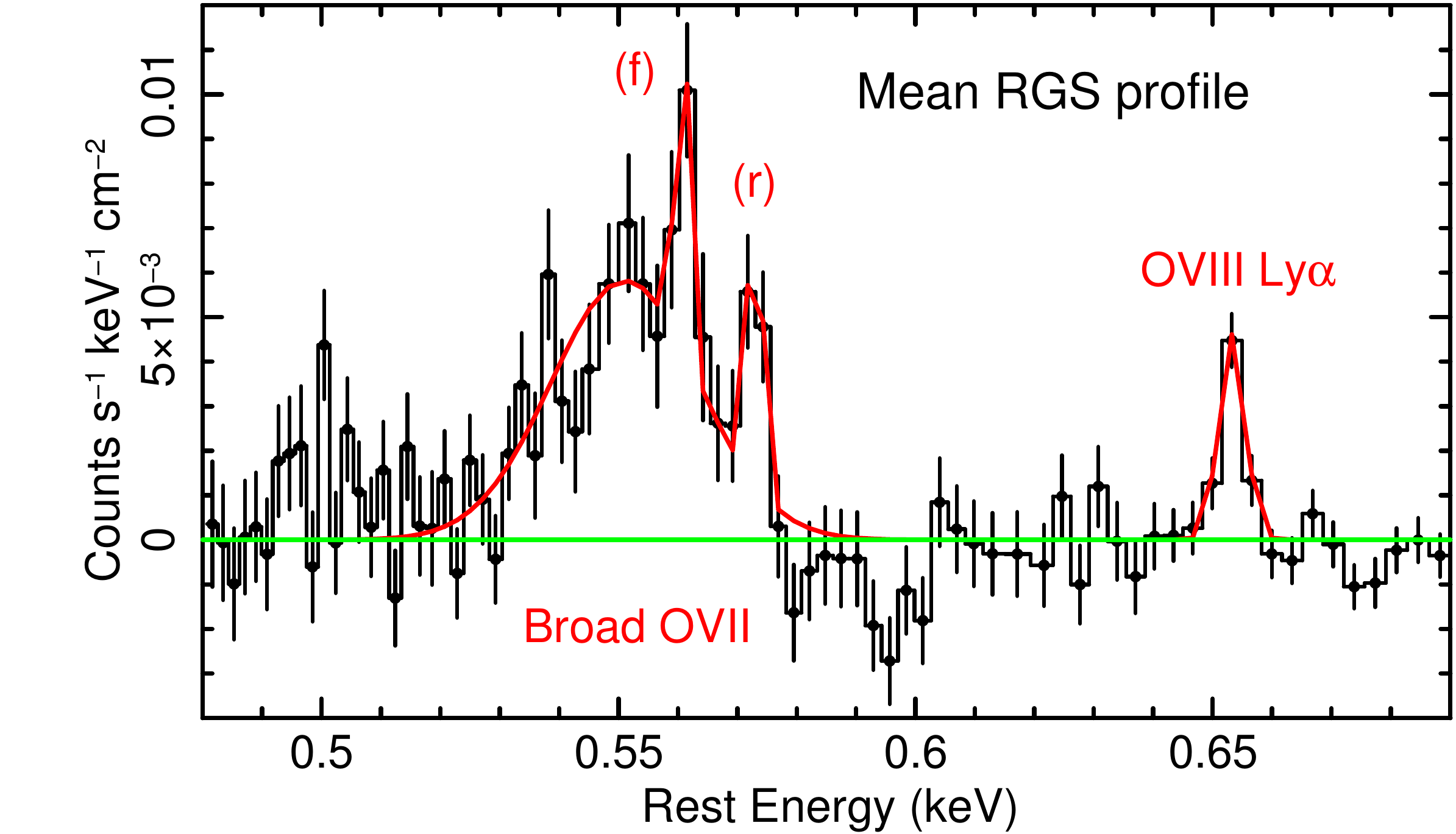}}
\rotatebox{-90}{\includegraphics[height=9cm]{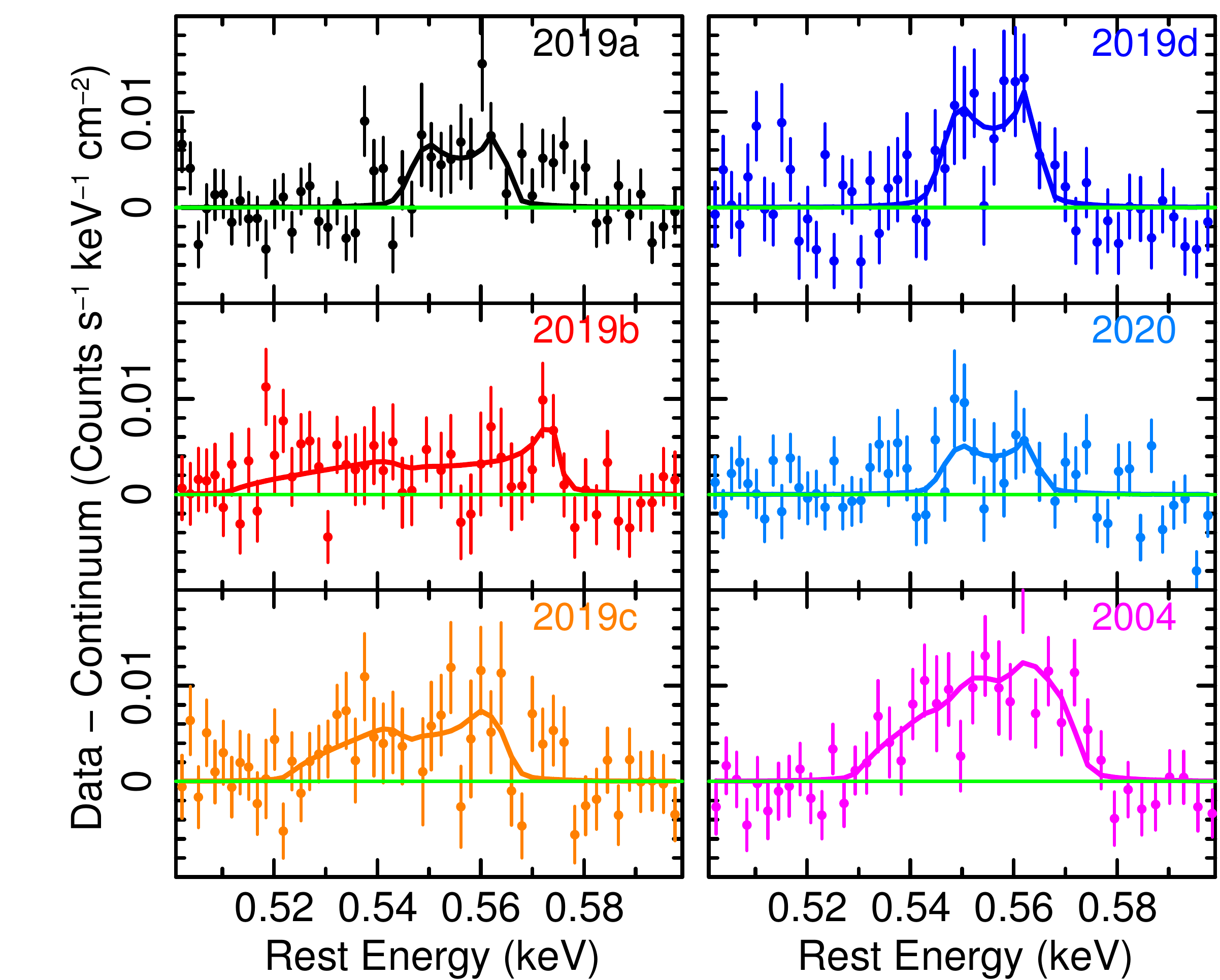}}
\end{center}
\caption{Data minus continuum residuals for the mean spectrum (upper panel) and each of the six RGS observations (lower six panels), over the O\,\textsc{vii} band. Overlaid (solid line) is the best-fit line profile, fitted with the \textsc{kyrline} model for the six observations and with the composite Gaussian model for the mean spectrum. The line flux is variable between the six spectra, being strongest at higher fluxes (2004 observation) and weakest at lower fluxes (2019a and 2020 observations). The profile shape is also variable, which may be attributed to arising from different disk radii, the broadest profiles (2004, 2019b, and c) occurring from emission closer to the black hole. The line parameters are listed in Table~\ref{tab:oxygen}.}
\label{fig:profiles}
\end{figure} 

The six spectra were fitted varying the disk inner radius ($R_{\rm in}$) 
and line flux (normalisation) between epochs. The inclination was also fitted for, but was assumed to be invariant across the datasets. 
The best-fit value is $\theta=9.9^{+1.0}_{-1.4}\degg$, consistent with a pole-on disk as is implied by the optical emission line profiles \citep{Kollatschny03, Liu17}. 
The resulting O\,\textsc{vii} profiles are plotted in Figure~\ref{fig:profiles} for each dataset, with the best-fit models overlaid and the fit results are reported in Table~\ref{tab:oxygen}. The profiles appear variable in both intensity and shape; for example, the original 2004 profile is highly broadened and intense, while the 2019c profile has a similar shape, but is of a lower intensity. The 2019d profile appears double-peaked, while the weakest profiles are found in the lowest flux 2019a and 2020 observations. The 2019b profile appears to be an outlier as the profile is very broad, but it has a low contrast against the continuum. The profile changes can be explained by a combination of variations in intensity and the inner radius, whereby broader profiles are produced for smaller $R_{\rm in}$ values. 
The overall fit-statistic obtained across the six spectra is $C/\nu=2156/2024$. The fit becomes significantly worse (by $\Delta C=40$ for $\Delta\nu=5$) if the inner radius is assumed to be invariant across the spectra and further still (by $\Delta C=34$ for $\Delta\nu=5$) if the line flux is also assumed to be constant. 

\subsection{A correlation between the line and continuum flux}

\begin{figure}
\begin{center}
\hspace{-0.5cm}
\rotatebox{-90}{\includegraphics[height=9cm]{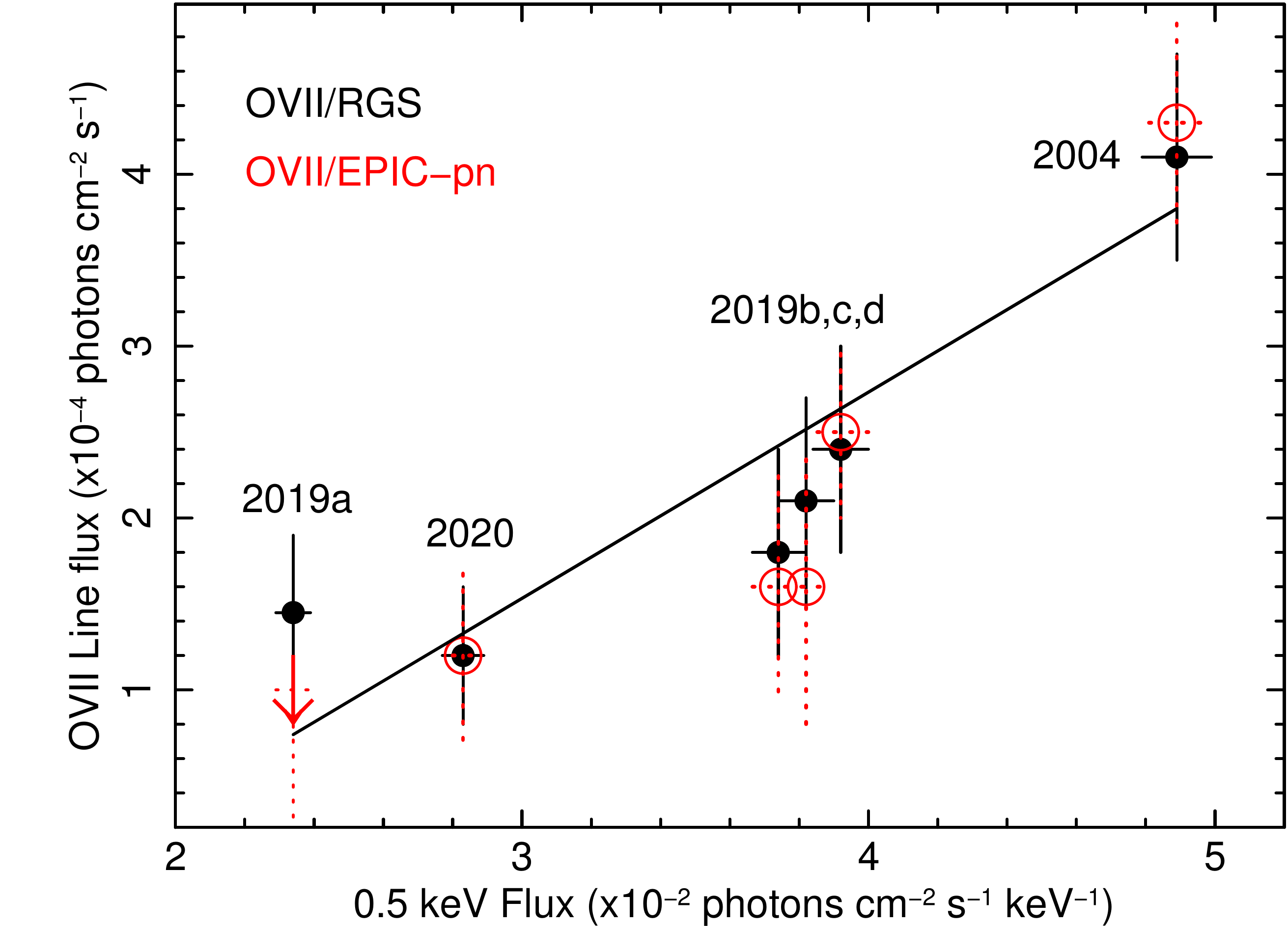}}
\end{center}
\caption{O\,\textsc{vii} line flux, as inferred by the \textsc{kyrline} model, plotted against the monochromatic continuum flux at 0.5\,keV. The six RGS observations show a positive trend, with the line flux (black circles) increasing with the continuum flux. The solid line shows the best fit correlation to the RGS data, which is significant at 99.7\% confidence. A positive trend is also inferred from the EPIC-pn data, the red open circles (plus an upper limit in 2019a, red arrow) show these corresponding line fluxes.}
\label{fig:correlation}
\end{figure}

Figure~\ref{fig:correlation} shows the RGS fluxes of the O\,\textsc{vii} lines, as derived from the \textsc{kyrline} model, plotted against the continuum flux at 0.5\,keV.  
The line flux appears to be correlated with the continuum, that is the strongest line occurs in the highest flux 2004 observation and the weakest are in the 
lowest flux spectra (2019a, 2020). The correlation was quantified by fitting a linear relation between the line and continuum flux, of the form $N_{\rm OVII} = a N_{\rm 0.5\,keV} + c$, where  
$N_{\rm OVII}$ is the line photon flux and $N_{\rm 0.5\,keV}$ is the monochromatic continuum flux at 0.5\,keV (in units of photons\,cm$^{-2}$\,s$^{-1}$\,keV$^{-1}$), as is reported in Table~\ref{tab:oxygen} and Figure~\ref{fig:correlation}. This gave a good fit ($\chi^2/\nu=3.8/4$), with a gradient of $a=0.98\pm0.26$ and an intercept of $c=-1.3\pm0.9$. A constant line flux returned a poor fit, with 
$\chi^2/\nu=18.3/5$, rejected at 99.7\% confidence. 

An analysis of the excess line flux in the six EPIC-pn spectra of Mrk\,110 (see Appendix~A) also confirms the positive correlation. 
The line fluxes, as modelled with a Gaussian line, were measured in 5/6 EPIC-pn spectra and are listed in Table~\ref{tab:oxygen}, along with one upper-limit (2019a). These are plotted in Figure~\ref{fig:correlation} (red points), alongside the RGS measurements and are consistent within errors. A linear correlation to the pn data-points returned a gradient of $a=1.35\pm0.29$ and intercept of $c=-2.8\pm1.1$, also consistent with the RGS. A constant line flux is ruled out at 99.98\% confidence (with $\chi^{2}/\nu=24.3/5$). 

The rms (root mean square) variability spectrum was computed across all six RGS observations to provide a further assessment of the variability of the O\,\textsc{vii} line. The methodology and results 
are described in detail in Appendix~B. In summary, an excess of counts was detected in the rms spectrum over the O\,\textsc{vii} band, compared to a variable powerlaw continuum. 
This also confirms the line variability and thus the positive correlation between the O\,\textsc{vii} line and continuum flux in Mrk\,110.

\section{Discussion and conclusions}

The analysis of six \xmm\ RGS observations of Mrk\,110 has confirmed the presence of the broad O\,\textsc{vii} line, as was first reported by \citet{Boller07}. Furthermore, we show that the line originates from a face-on accretion disk, which is consistent with the orientation of the optical BLR \citep{Kollatschny03,Liu17}, and it appears to respond to the continuum flux. 
Notably, relativistic disk lines from H-like O, N, and C were first claimed in the RGS spectra of the Seyfert 1 galaxies, \object{MCG\,$-$6-30-15} and \object{Mrk\,766} \citep{BR01, Sako03}, as well as in a small number of other AGN, including, for example, \object{NGC 4051} \citep{Ogle04} and \object{ESO\,198$-$G24} \citep{Porquet04}. The former two cases have since been disputed, with the features possibly arising from a dusty warm absorber \citep{Lee01, Turner03}, while the broad O profiles in NGC\,4051 are consistent with the optical-UV broad line region \citep{Peretz19}. In Mrk\,110, there is no detectable warm absorption and the broad oxygen emission is revealed against the otherwise featureless, bare continuum. 

The presence of broad O\,\textsc{vii} or O\,\textsc{viii} lines originating from BLR gas appears to be frequent in AGN X-ray spectra \citep{Costantini07, Longinotti08, Longinotti10, Detmers11, Reeves13, Reeves16}. However, in Mrk 110, the O\,\textsc{vii} line width is significantly broader than for the optical broad lines. The velocity width inferred from the mean RGS profile is $v_{\rm FWHM}=15900\pm1800$\,km\,s$^{-1}$, while \citet{Kollatschny03} measured widths of $v_{\rm FWHM}=4444\pm200$\,km\,s$^{-1}$ (for He\,\textsc{ii} $\lambda4686$) and $v_{\rm FWHM}=1510\pm100$\,km\,s$^{-1}$ (for H$\beta$). 
The radial location of the optical BLR emission in Mrk\,110 lies between $10^{16}-10^{17}$\,cm, as inferred from the lags between the optical lines and the continuum of between $4-32$\,days \citep{Kollatschny03}. 
The broad O\,\textsc{vii} emission region likely resides inside this, given its greater line width, at tens of gravitational radii as inferred by the disk line modelling. 
This is consistent with the gravitational redshift of the O\,\textsc{vii} profile of $\Delta z\sim 0.02-0.03$, as inferred both from the mean line energy obtained from these observations and the line energy computed 
in \citet{Boller07} for the 2004 observation. 

To determine the physical properties of the X-ray line emitting gas, the Mrk\,110 RGS spectra were modelled with the photoionised emission predicted by \textsc{xstar}, 
convolved with the velocity broadening from the accretion disk.  
Fitting the 2004 RGS spectrum resulted in an ionisation parameter of $\log(\xi/{\rm erg\,cm\,s^{-1}})=1.2\pm0.3$, where the dominance of the O\,\textsc{vii} emission over O\,\textsc{viii} sets the gas ionisation. 
The gas density was calculated via $n_{\rm e}=L_{\rm ion}/\xi R^{2}$, where $L_{\rm ion}=10^{45}$\,erg\,s$^{-1}$ is the 1--1000\,Rydberg ionising luminosity for Mrk\,110. At a distance of $R=5\times10^{14}\,{\rm cm}\sim30R_{\rm g}$, the density is then $n_{\rm e}=3\times10^{14}$\,cm$^{-3}$. This is much higher than expected for BLR clouds. 

The density of a standard $\alpha$ accretion disk was computed in the case where the inner region is radiation dominated. From equation 1 in \citet{Garcia16} (also see \citealt{Svensson94}) and for a black hole mass of $M_{\rm BH}=10^{8}$\,M$_{\odot}$ with a viscosity parameter of $\alpha=0.1$, then the predicted density versus radius is as follows:
\[ n_{\rm e} \approx 7\times10^{12} \dot{m}^{-2} r^{3/2} \left[1-\left(\frac{3}{r}\right)\right]^{-1} (1-f)^{-3}\,{\rm cm}^{-3} \]
\noindent where $r$ is in Schwarzschild units.  Here $\dot{m}$ is the dimensionless accretion rate, defined as $\dot{m}=\dot{M} / (\eta \dot{M}_{\rm Edd}) = \dot{M}c^2 (L_{\rm Edd})^{-1}$, while $f$ is the fraction of power radiated in the corona. In Mrk\,110, for an efficiency of $\eta=0.1$ and an Eddington ratio of $\sim10$\%, then $\dot{m}$ is near unity. Thus for $f\sim0.1$ and a similar radius above ($r\sim15R_{\rm S}$), the disk density is predicted to be of the order of a few times $10^{14}$\,cm$^{-3}$. This is consistent with the density derived above and an origin in the accretion disk appears plausible. 

The broad O\,\textsc{vii} line is variable, on timescales down to 2 days between observations. The line flux appears to respond to the soft X-ray continuum level, as is shown in Figure~\ref{fig:correlation}. This is consistent with the short timescale delays ($\tau=4\pm2$\,days) seen for the He\,\textsc{ii} line, while the broadest (FWHM of 6000\,km\,s$^{-1}$) component of He\,\textsc{ii} also shows pronounced variability by up to a factor of eight in flux \citep{Veron-cetty07}. The O\,\textsc{vii} behaviour may be an extension of this, but from where this higher excitation line originates closer to the black hole in a face-on system. 

Similar to the He\,\textsc{ii} line, the broad O\,\textsc{vii} line appears weak and relatively narrow (or from larger radii, Table~\ref{tab:oxygen}) in the lowest flux spectra, as seen in the 2019a and 2020 observations. From Figure~\ref{fig:swiftzoom}, the 2019a observation occurs during a dip in the light curve, while the April 2020 observation occurs prior to a prolonged (days-long) flare in the X-ray light curve. The strongest broad profile occurs in the highest flux 2004 observation, but which did not coincide with any X-ray monitoring. However the behaviour is similar to what has been recently measured in the Fe K lines of some AGN, where the broad disk components 
emerge only during brighter episodes, for example in Ark\,120 \citep{Porquet18}, \object{NGC\,2992} \citep{Shu10, Marinucci18, Marinucci20}, and \object{PDS 456} \citep{Reeves21}.  
Here, the O\,\textsc{vii} variability could probe reverberation in the inner disk, where an increase in line flux and width might proceed strong flares in the X-ray light curve. Future, more intensive monitoring of Mrk\,110, with \xmm\ RGS and {\it Swift}, could reveal this behaviour.

\begin{acknowledgements}

We thank the referee for their helpful suggestions. The paper is based on observations obtained with the {\sl XMM-Newton}, and ESA science
mission with instruments and contributions directly funded by ESA
member states and the USA (NASA). D.P. and N.G. gratefully acknowledge financial support from the Centre National d'Etudes Spatiales (CNES).
\end{acknowledgements}


\bibliographystyle{aa}
\bibliography{biblio}

\begin{thebibliography}{48}
\expandafter\ifx\csname natexlab\endcsname\relax\def\natexlab#1{#1}\fi

\bibitem[{{Bischoff} \& {Kollatschny}(1999)}]{Bischoff99}
{Bischoff}, K. \& {Kollatschny}, W. 1999, \aap, 345, 49

\bibitem[{{Boller} {et~al.}(2007){Boller}, {Balestra}, \&
  {Kollatschny}}]{Boller07}
{Boller}, T., {Balestra}, I., \& {Kollatschny}, W. 2007, \aap, 465, 87

\bibitem[{{Boroson} \& {Green}(1992)}]{Boroson92}
{Boroson}, T.~A. \& {Green}, R.~F. 1992, \apjs, 80, 109

\bibitem[{{Branduardi-Raymont} {et~al.}(2001){Branduardi-Raymont}, {Sako},
  {Kahn}, {Brinkman}, {Kaastra}, \& {Page}}]{BR01}
{Branduardi-Raymont}, G., {Sako}, M., {Kahn}, S.~M., {et~al.} 2001, \aap, 365,
  L140

\bibitem[{{Cash}(1979)}]{Cash79}
{Cash}, W. 1979, \apj, 228, 939

\bibitem[{{Costantini} {et~al.}(2007){Costantini}, {Kaastra}, {Arav}, {Kriss},
  {Steenbrugge}, {Gabel}, {Verbunt}, {Behar}, {Gaskell}, {Korista}, {Proga},
  {Quijano}, {Scott}, {Klimek}, \& {Hedrick}}]{Costantini07}
{Costantini}, E., {Kaastra}, J.~S., {Arav}, N., {et~al.} 2007, \aap, 461, 121

\bibitem[{{Czerny} \& {Elvis}(1987)}]{Czerny87}
{Czerny}, B. \& {Elvis}, M. 1987, \apj, 321, 305

\bibitem[{{Dauser} {et~al.}(2010){Dauser}, {Wilms}, {Reynolds}, \&
  {Brenneman}}]{Dauser10}
{Dauser}, T., {Wilms}, J., {Reynolds}, C.~S., \& {Brenneman}, L.~W. 2010,
  \mnras, 409, 1534

\bibitem[{{den Herder} {et~al.}(2001){den Herder}, {Brinkman}, {Kahn}, {Brand
  uardi-Raymont}, {Thomsen}, {Aarts}, {Audard}, {Bixler}, {den Boggende},
  {Cottam}, {Decker}, {Dubbeldam}, {Erd}, {Goulooze}, {G{\"u}del}, {Guttridge},
  {Hailey}, {Janabi}, {Kaastra}, {de Korte}, {van Leeuwen}, {Mauche},
  {McCalden}, {Mewe}, {Naber}, {Paerels}, {Peterson}, {Rasmussen}, {Rees},
  {Sakelliou}, {Sako}, {Spodek}, {Stern}, {Tamura}, {Tandy}, {de Vries},
  {Welch}, \& {Zehnder}}]{denHerder01}
{den Herder}, J.~W., {Brinkman}, A.~C., {Kahn}, S.~M., {et~al.} 2001, \aap,
  365, L7

\bibitem[{{Detmers} {et~al.}(2011){Detmers}, {Kaastra}, {Steenbrugge},
  {Ebrero}, {Kriss}, {Arav}, {Behar}, {Costantini}, {Branduardi-Raymont},
  {Mehdipour}, {Bianchi}, {Cappi}, {Petrucci}, {Ponti}, {Pinto}, {Ratti}, \&
  {Holczer}}]{Detmers11}
{Detmers}, R.~G., {Kaastra}, J.~S., {Steenbrugge}, K.~C., {et~al.} 2011, \aap,
  534, A38

\bibitem[{{Dov{\v c}iak} {et~al.}(2004){Dov{\v c}iak}, {Karas}, \&
  {Yaqoob}}]{Dovciak04}
{Dov{\v c}iak}, M., {Karas}, V., \& {Yaqoob}, T. 2004, \apjs, 153, 205

\bibitem[{{Emmanoulopoulos} {et~al.}(2011){Emmanoulopoulos}, {Papadakis},
  {McHardy}, {Nicastro}, {Bianchi}, \& {Ar{\'e}valo}}]{Emman11}
{Emmanoulopoulos}, D., {Papadakis}, I.~E., {McHardy}, I.~M., {et~al.} 2011,
  \mnras, 415, 1895

\bibitem[{{Garc{\'{\i}}a} {et~al.}(2016){Garc{\'{\i}}a}, {Fabian}, {Kallman},
  {Dauser}, {Parker}, {McClintock}, {Steiner}, \& {Wilms}}]{Garcia16}
{Garc{\'{\i}}a}, J.~A., {Fabian}, A.~C., {Kallman}, T.~R., {et~al.} 2016,
  \mnras, 462, 751

\bibitem[{{Gierli{\'n}ski} \& {Done}(2004)}]{Gierlinski04}
{Gierli{\'n}ski}, M. \& {Done}, C. 2004, \mnras, 349, L7

\bibitem[{{Grupe}(2004)}]{Grupe04}
{Grupe}, D. 2004, \aj, 127, 1799

\bibitem[{{Kalberla} {et~al.}(2005){Kalberla}, {Burton}, {Hartmann}, {Arnal},
  {Bajaja}, {Morras}, \& {P{\"o}ppel}}]{Kalberla05}
{Kalberla}, P.~M.~W., {Burton}, W.~B., {Hartmann}, D., {et~al.} 2005, \aap,
  440, 775

\bibitem[{{Kinkhabwala} {et~al.}(2002){Kinkhabwala}, {Sako}, {Behar}, {Kahn},
  {Paerels}, {Brinkman}, {Kaastra}, {Gu}, \& {Liedahl}}]{Kinkhabwala02}
{Kinkhabwala}, A., {Sako}, M., {Behar}, E., {et~al.} 2002, \apj, 575, 732

\bibitem[{{Kollatschny}(2003)}]{Kollatschny03}
{Kollatschny}, W. 2003, \aap, 412, L61

\bibitem[{{Lee} {et~al.}(2001){Lee}, {Ogle}, {Canizares}, {Marshall}, {Schulz},
  {Morales}, {Fabian}, \& {Iwasawa}}]{Lee01}
{Lee}, J.~C., {Ogle}, P.~M., {Canizares}, C.~R., {et~al.} 2001, \apjl, 554, L13

\bibitem[{{Liu} {et~al.}(2017){Liu}, {Feng}, \& {Bai}}]{Liu17}
{Liu}, H.~T., {Feng}, H.~C., \& {Bai}, J.~M. 2017, \mnras, 466, 3323

\bibitem[{{Lohfink} {et~al.}(2016){Lohfink}, {Reynolds}, {Pinto}, {Alston},
  {Boggs}, {Christensen}, {Craig}, {Fabian}, {Hailey}, {Harrison}, {Kara},
  {Matt}, {Parker}, {Stern}, {Walton}, \& {Zhang}}]{Lohfink16}
{Lohfink}, A.~M., {Reynolds}, C.~S., {Pinto}, C., {et~al.} 2016, \apj, 821, 11

\bibitem[{{Longinotti} {et~al.}(2010){Longinotti}, {Costantini}, {Petrucci},
  {Boisson}, {Mouchet}, {Santos-Lleo}, {Matt}, {Ponti}, \&
  {Gon{\c{c}}alves}}]{Longinotti10}
{Longinotti}, A.~L., {Costantini}, E., {Petrucci}, P.~O., {et~al.} 2010, \aap,
  510, A92

\bibitem[{{Longinotti} {et~al.}(2008){Longinotti}, {Nucita}, {Santos-Lleo}, \&
  {Guainazzi}}]{Longinotti08}
{Longinotti}, A.~L., {Nucita}, A., {Santos-Lleo}, M., \& {Guainazzi}, M. 2008,
  \aap, 484, 311

\bibitem[{{Marinucci} {et~al.}(2020){Marinucci}, {Bianchi}, {Braito}, {De
  Marco}, {Matt}, {Middei}, {Nardini}, \& {Reeves}}]{Marinucci20}
{Marinucci}, A., {Bianchi}, S., {Braito}, V., {et~al.} 2020, \mnras, 496, 3412

\bibitem[{{Marinucci} {et~al.}(2018){Marinucci}, {Bianchi}, {Braito}, {Matt},
  {Nardini}, \& {Reeves}}]{Marinucci18}
{Marinucci}, A., {Bianchi}, S., {Braito}, V., {et~al.} 2018, \mnras, 478, 5638

\bibitem[{{Miller} {et~al.}(2009){Miller}, {Turner}, \& {Reeves}}]{Miller09}
{Miller}, L., {Turner}, T.~J., \& {Reeves}, J.~N. 2009, \mnras, 399, L69

\bibitem[{{Nandra} {et~al.}(2007){Nandra}, {O'Neill}, {George}, \&
  {Reeves}}]{Nandra07}
{Nandra}, K., {O'Neill}, P.~M., {George}, I.~M., \& {Reeves}, J.~N. 2007,
  \mnras, 382, 194

\bibitem[{{Ogle} {et~al.}(2004){Ogle}, {Mason}, {Page}, {Salvi}, {Cordova},
  {McHardy}, \& {Priedhorsky}}]{Ogle04}
{Ogle}, P.~M., {Mason}, K.~O., {Page}, M.~J., {et~al.} 2004, \apj, 606, 151

\bibitem[{{Patrick} {et~al.}(2012){Patrick}, {Reeves}, {Porquet}, {Markowitz},
  {Braito}, \& {Lobban}}]{Patrick12}
{Patrick}, A.~R., {Reeves}, J.~N., {Porquet}, D., {et~al.} 2012, \mnras, 426,
  2522

\bibitem[{{Peretz} {et~al.}(2019){Peretz}, {Miller}, \& {Behar}}]{Peretz19}
{Peretz}, U., {Miller}, J.~M., \& {Behar}, E. 2019, \apj, 879, 102

\bibitem[{{Porquet} \& {Dubau}(2000)}]{Porquet00}
{Porquet}, D. \& {Dubau}, J. 2000, \aaps, 143, 495

\bibitem[{{Porquet} {et~al.}(2004){Porquet}, {Kaastra}, {Page}, {O'Brien},
  {Ward}, \& {Dubau}}]{Porquet04}
{Porquet}, D., {Kaastra}, J.~S., {Page}, K.~L., {et~al.} 2004, \aap, 413, 913

\bibitem[{{Porquet} {et~al.}(2018){Porquet}, {Reeves}, {Matt}, {Marinucci},
  {Nardini}, {Braito}, {Lobban}, {Ballantyne}, {Boggs}, {Christensen},
  {Dauser}, {Farrah}, {Garcia}, {Hailey}, {Harrison}, {Stern}, {Tortosa},
  {Ursini}, \& {Zhang}}]{Porquet18}
{Porquet}, D., {Reeves}, J.~N., {Matt}, G., {et~al.} 2018, \aap, 609, A42

\bibitem[{{Reeves} {et~al.}(2021){Reeves}, {Braito}, {Porquet}, {Lobban},
  {Matzeu}, \& {Nardini}}]{Reeves21}
{Reeves}, J.~N., {Braito}, V., {Porquet}, D., {et~al.} 2021, \mnras, 500, 1974

\bibitem[{{Reeves} {et~al.}(2013){Reeves}, {Porquet}, {Braito}, {Gofford},
  {Nardini}, {Turner}, {Crenshaw}, \& {Kraemer}}]{Reeves13}
{Reeves}, J.~N., {Porquet}, D., {Braito}, V., {et~al.} 2013, \apj, 776, 99

\bibitem[{{Reeves} {et~al.}(2016){Reeves}, {Porquet}, {Braito}, {Nardini},
  {Lobban}, \& {Turner}}]{Reeves16}
{Reeves}, J.~N., {Porquet}, D., {Braito}, V., {et~al.} 2016, \apj, 828, 98

\bibitem[{{Risaliti} {et~al.}(2013){Risaliti}, {Harrison}, {Madsen}, {Walton},
  {Boggs}, {Christensen}, {Craig}, {Grefenstette}, {Hailey}, {Nardini},
  {Stern}, \& {Zhang}}]{Risaliti13}
{Risaliti}, G., {Harrison}, F.~A., {Madsen}, K.~K., {et~al.} 2013, \nat, 494,
  449

\bibitem[{{Sako} {et~al.}(2003){Sako}, {Kahn}, {Branduardi-Raymont}, {Kaastra},
  {Brinkman}, {Page}, {Behar}, {Paerels}, {Kinkhabwala}, {Liedahl}, \& {den
  Herder}}]{Sako03}
{Sako}, M., {Kahn}, S.~M., {Branduardi-Raymont}, G., {et~al.} 2003, \apj, 596,
  114

\bibitem[{{Shu} {et~al.}(2010){Shu}, {Yaqoob}, {Murphy}, {Braito}, {Wang}, \&
  {Zheng}}]{Shu10}
{Shu}, X.~W., {Yaqoob}, T., {Murphy}, K.~D., {et~al.} 2010, \apj, 713, 1256

\bibitem[{{Str{\"u}der} {et~al.}(2001){Str{\"u}der}, {Briel}, {Dennerl},
  {Hartmann}, {Kendziorra}, {Meidinger}, {Pfeffermann}, {Reppin}, {Aschenbach},
  {Bornemann}, {Br{\"a}uninger}, {Burkert}, {Elender}, {Freyberg}, {Haberl},
  {Hartner}, {Heuschmann}, {Hippmann}, {Kastelic}, {Kemmer}, {Kettenring},
  {Kink}, {Krause}, {M{\"u}ller}, {Oppitz}, {Pietsch}, {Popp}, {Predehl},
  {Read}, {Stephan}, {St{\"o}tter}, {Tr{\"u}mper}, {Holl}, {Kemmer}, {Soltau},
  {St{\"o}tter}, {Weber}, {Weichert}, {von Zanthier}, {Carathanassis}, {Lutz},
  {Richter}, {Solc}, {B{\"o}ttcher}, {Kuster}, {Staubert}, {Abbey}, {Holland},
  {Turner}, {Balasini}, {Bignami}, {La Palombara}, {Villa}, {Buttler},
  {Gianini}, {Lain{\'e}}, {Lumb}, \& {Dhez}}]{Struder01}
{Str{\"u}der}, L., {Briel}, U., {Dennerl}, K., {et~al.} 2001, \aap, 365, L18

\bibitem[{{Svensson} \& {Zdziarski}(1994)}]{Svensson94}
{Svensson}, R. \& {Zdziarski}, A.~A. 1994, \apj, 436, 599

\bibitem[{{Titarchuk}(1994)}]{Titarchuk94}
{Titarchuk}, L. 1994, \apj, 434, 570

\bibitem[{{Turner} {et~al.}(2003){Turner}, {Fabian}, {Vaughan}, \&
  {Lee}}]{Turner03}
{Turner}, A.~K., {Fabian}, A.~C., {Vaughan}, S., \& {Lee}, J.~C. 2003, \mnras,
  346, 833

\bibitem[{{Vaughan} {et~al.}(2003){Vaughan}, {Edelson}, {Warwick}, \&
  {Uttley}}]{Vaughan03}
{Vaughan}, S., {Edelson}, R., {Warwick}, R.~S., \& {Uttley}, P. 2003, \mnras,
  345, 1271

\bibitem[{{Vaughan} {et~al.}(2004){Vaughan}, {Fabian}, {Ballantyne}, {De Rosa},
  {Piro}, \& {Matt}}]{Vaughan04}
{Vaughan}, S., {Fabian}, A.~C., {Ballantyne}, D.~R., {et~al.} 2004, \mnras,
  351, 193

\bibitem[{{Verner} {et~al.}(1996){Verner}, {Ferland}, {Korista}, \&
  {Yakovlev}}]{Verner96}
{Verner}, D.~A., {Ferland}, G.~J., {Korista}, K.~T., \& {Yakovlev}, D.~G. 1996,
  \apj, 465, 487

\bibitem[{{V{\'e}ron-Cetty} {et~al.}(2007){V{\'e}ron-Cetty}, {V{\'e}ron},
  {Joly}, \& {Kollatschny}}]{Veron-cetty07}
{V{\'e}ron-Cetty}, M.~P., {V{\'e}ron}, P., {Joly}, M., \& {Kollatschny}, W.
  2007, \aap, 475, 487

\bibitem[{{Wilms} {et~al.}(2000){Wilms}, {Allen}, \& {McCray}}]{Wilms00}
{Wilms}, J., {Allen}, A., \& {McCray}, R. 2000, \apj, 542, 914

\end{thebibliography}

\begin{appendix}

\section{Analysis of the EPIC-pn data}

\subsection{Data reduction}

\begin{figure*}[ht]
\begin{center}
\rotatebox{0}{\includegraphics[width=14.5cm]{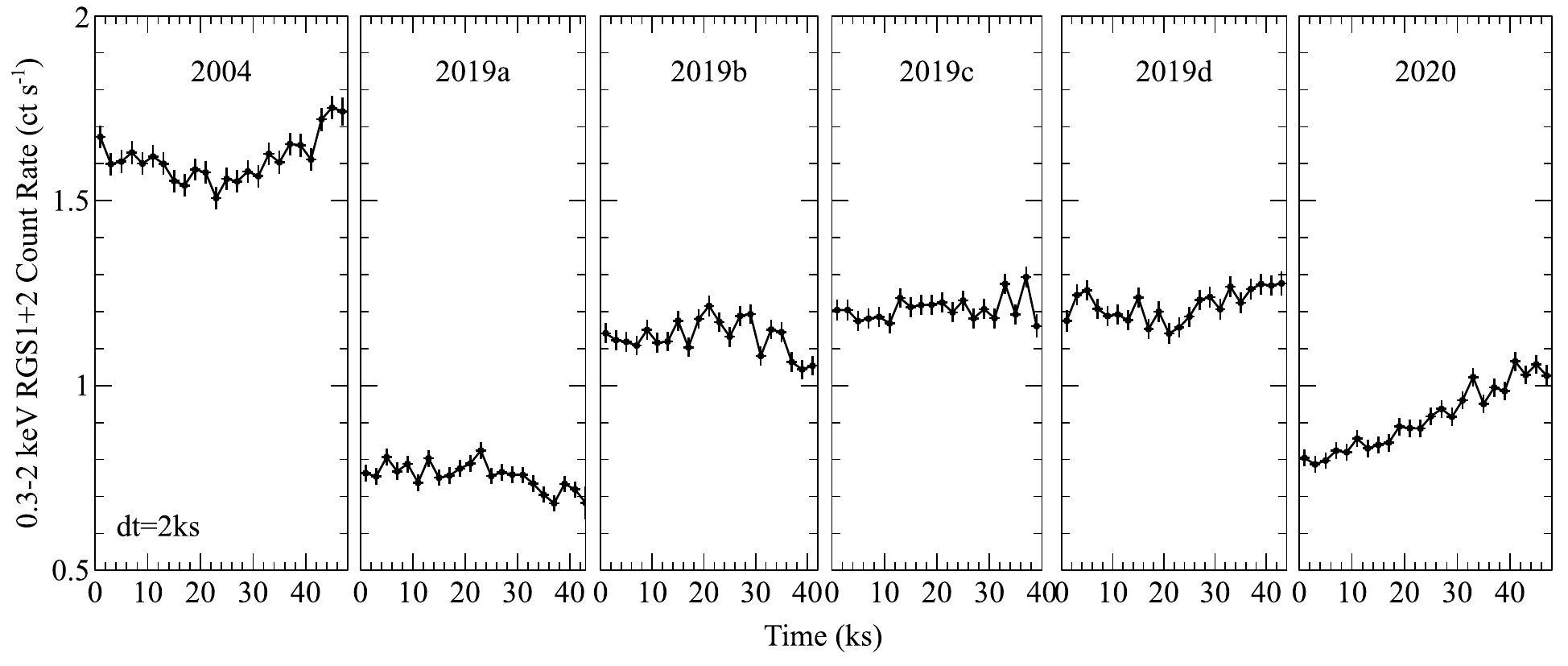}}
\end{center}
\vspace*{-0.5cm}
\caption{\xmm\ RGS\,1+2 light curve of all six observations, extracted over the 0.3--2.0\,keV band. Time is in kiloseconds measured from the start of each exposure. 
2004 is the brightest observation and 2019a and 2020 are the faintest ones, with 2019b--d being at a similar flux level.}
\label{fig:rgslc}
\end{figure*} 

The XMM-Newton EPIC (European Photon Imaging Camera) event files were processed with the Science Analysis System (SAS) v18.0.0, applying the latest available calibration. 
Due to the high source brightness, all the EPIC instruments were operated in small window (SW) mode. 
However, this was not sufficient to prevent photon pile-up in
the MOS cameras and therefore only the EPIC-pn \citep{Struder01} data 
were analysed. The EPIC-pn spectra were extracted using event patterns 0--4, that is from single and double pixel events.  
Source spectra were obtained from a circular region centred on Mrk\,110, with a radius of
35${\arcsec}$, to avoid the edge of the chip. The background spectra were extracted from a rectangular region in the lower part of
the CCD window that contains no (or negligible) source photons.  
Redistribution matrices and ancillary response files for the six pn spectra were generated with the SAS tasks
{\sc rmfgen} and {\sc arfgen}. The subsequent source spectra were binned to a minimum of 100 counts per bin and  
$\chi^2$ minimalisation was employed for spectral fitting. The background rates are less than 1\% of the net source rates over the 0.3--10\,keV band. 

The total net exposures (about 30\,ks per observation, accounting for CCD deadtime) and the 0.3--10\,keV band count rates 
are listed in Table~\ref{tab:pn}. Three of the observations, 2019a--c, were performed with the thick optical filter in place in front of the EPIC-pn camera, while the remaining three 
used the thin filter. The result of this is that the count rates in the thick filter observations were attenuated in the soft band by up to about 40\% at 0.5\,keV, compared to the 
thin filter observations.  
This was accounted for in the ancillary response files, via the reduction of the pn effective area at soft X-rays in the effective area curve. 
The 0.3--2.0\,keV count rate light curves of the six Mrk\,110 observation are shown in Figure~\ref{fig:rgslc}. As the EPIC-pn observations employed different optical filters, the RGS\,1+2 
light curves are plotted instead to provide a like-for-like comparison between the six observations. As is discussed in the main article, the 2019a observation is the faintest one by about a factor of two compared to the brightest 2004 observation, while the 2019b--d observations are of a similar soft X-ray flux. 

\begin{table*}[t!]
\caption{Parameters from the six EPIC-pn spectra.}
\begin{center}
\begin{tabular}{lccccccccc}
\hline \hline
OBS & Exp\,(ks) & Rate\,(s$^{-1}$) & $N_{\rm OVII}$$^a$ & EW$_{\rm OVII}^b$ & $\Delta \chi^2$$^c$ & $N_{\rm Fe}^{d}$ & EW$_{\rm Fe}^{b}$ & $F_{0.3-2\,{\rm keV}}^e$ & $F_{2-10\,{\rm keV}}^e$ \\
\hline
2004 & 32.9 & $22.91\pm0.03$ & $4.3\pm0.6$ & $9.3\pm1.3$ & 148.3 & $1.53\pm0.42$ & $47\pm13$ & $3.4\pm0.1$ & $3.0\pm0.1$ \\
2019a & 29.0 & $10.95\pm0.02$ & $<1.2$ & $<3.6$ & -- & $1.44\pm0.40$ & $60\pm17$ & $1.9\pm0.1$ & $2.1\pm0.1$ \\
2019b & 28.8 & $15.65\pm0.02$ & $1.6\pm0.8$ & $3.8\pm1.9$ & 11.7 & $1.18\pm0.44$ & $39\pm15$ & $2.8\pm0.1$ & $2.7\pm0.1$ \\
2019c & 27.1 & $16.72\pm0.02$ & $1.6\pm0.8$ & $3.6\pm1.8$ & 10.6 & $1.39\pm0.46$ & $45\pm15$ & $3.0\pm0.1$ & $2.8\pm0.1$ \\
2019d & 29.9 & $21.10\pm0.03$ & $2.5\pm0.6$ & $5.5\pm1.4$ & 44.7 & $1.63\pm0.45$ & $54\pm15$ & $2.9\pm0.1$ & $2.7\pm0.1$ \\
2020 & 32.7 & $16.50\pm0.02$ & $1.2\pm0.5$ & $3.7\pm1.5$ & 16.3 & $1.10\pm0.39$ & $41\pm15$ & $2.2\pm0.1$ & $2.3\pm0.1$ \\
\hline
\hline
\end{tabular}
\end{center}
\label{tab:pn}
\small{\textit{Notes.} $^{a}$Flux of the O\,\textsc{vii} line, fitted with a Gaussian profile. Units are $\times10^{-4}$\,photons\,cm$^{-2}$\,s$^{-1}$. 
$^{b}$Equivalent width of line in units of eV.
$^c$Improvement in $\chi^2$ upon adding the O\,\textsc{vii} line component.
$^{d}$Flux of the Fe K line, fitted with a Gaussian profile. Units are $\times10^{-5}$\,photons\,cm$^{-2}$\,s$^{-1}$.
$^{e}$Continuum flux from 0.3--2.0\,keV or 2--10\,keV in units $\times10^{-11}$\,ergs\,cm$^{-2}$\,s$^{-1}$. }
\end{table*}

\subsection{Spectral analysis}

\begin{figure}[ht]
\begin{center}
\hspace{-0.6cm}
\rotatebox{-90}{\includegraphics[height=9cm]{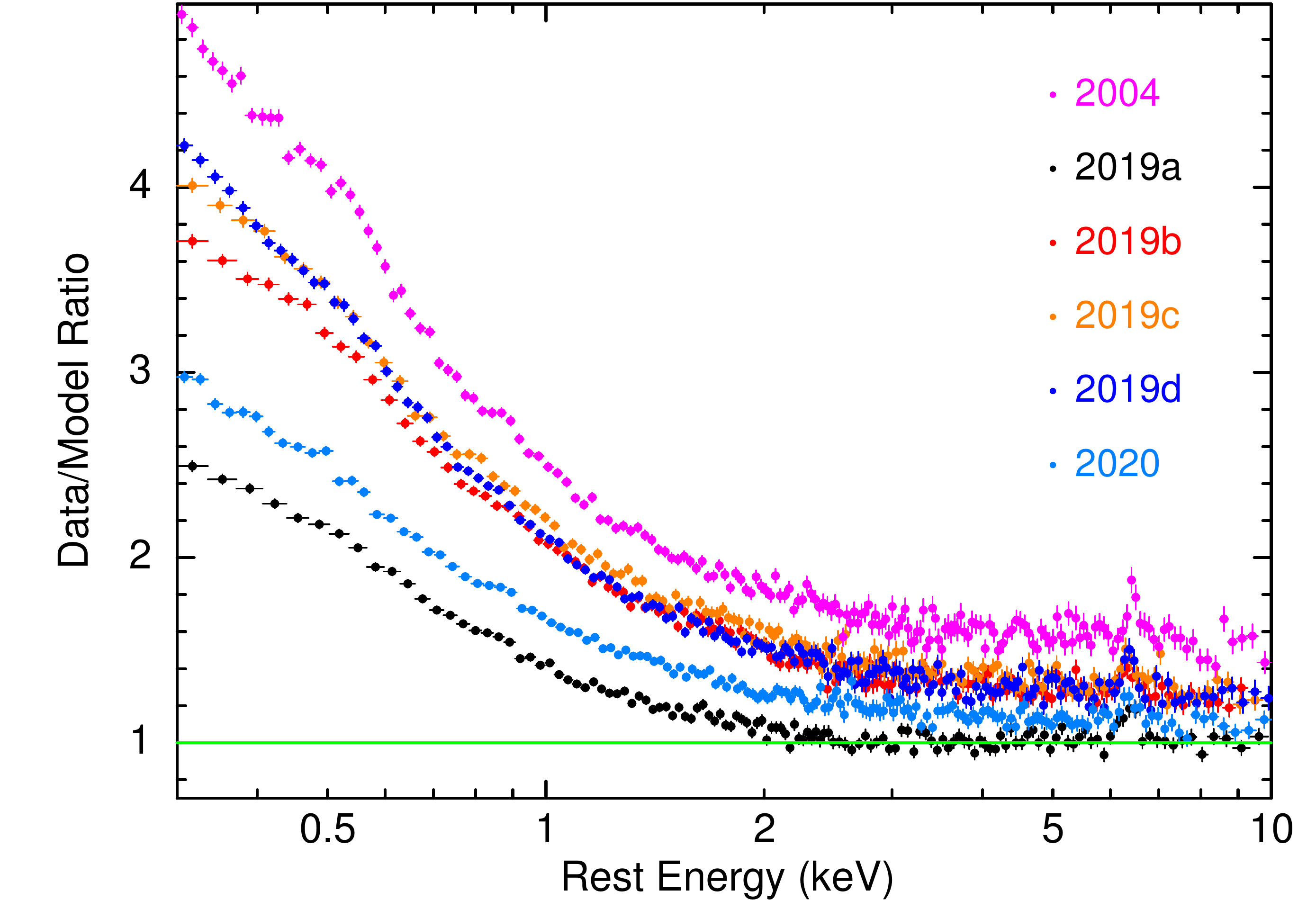}}
\end{center}
\caption{All six EPIC-pn spectra of Mrk\,110. The spectra are plotted as a ratio against a power continuum of $\Gamma=1.63\pm0.01$, fitted to the faintest 2019a observation over the 2.5--10\,keV band. All the spectra show a soft X-ray excess against the power-law continuum, with the excess being most prominent in the brightest observations. The continuum in all six spectra 
can be fitted with a two-component model, consisting of Comptonised soft X-ray emission from the disk and a hard X-ray power-law. A weak iron K line at 6.4\,keV is also present 
in all the observations. The EPIC-pn spectra have been re-binned by a further factor of $\times 5$ above the native binning of 100 counts per bin for display purposes.}
\label{fig:pnbroad}
\end{figure}

As there was no significant spectral variability within each of the individual observations, all six observations were subsequently fitted simultaneously. Compared to a simple power-law fitted between 2.5--10\,keV, all six observations show a prominent soft X-ray excess towards lower energies (Figure~\ref{fig:pnbroad}).  This soft excess is stronger in the higher flux observations 
and the spectra become slightly softer with increasing X-ray flux as a result. There is no evidence for any excess X-ray absorption in the spectra, aside from the Galactic column. 
The six spectra were subsequently fitted over the 0.3--10\,keV band, using a two component continuum model consisting of a hard X-ray power-law and thermal Comptonisation emission via the \textsc{comptt} model \citep{Titarchuk94}. The latter accounts for the soft excess seen in the spectra. The relative normalisations of the power-law and \textsc{comptt} 
components were allowed to vary between the observations.  The Comptonisation component has a best-fit electron temperature of $kT=0.44\pm0.04$\,keV and a high optical depth of 
$\tau=9.9\pm0.5$, while an input seed photon temperature of $kT=10$\,eV was adopted, which represents the tail of the big blue-bump emission from the disk.  
The power-law photon index varies slightly between $\Gamma=1.63\pm0.01$ to $\Gamma=1.73\pm0.01$. 
The X-ray luminosity of Mrk\,110 over the 0.3--10\,keV band varied between $1.2-1.9\times10^{44}$\,erg\,s$^{-1}$. 

\begin{figure*}[ht]
\begin{center}
\rotatebox{-90}{\includegraphics[height=9cm]{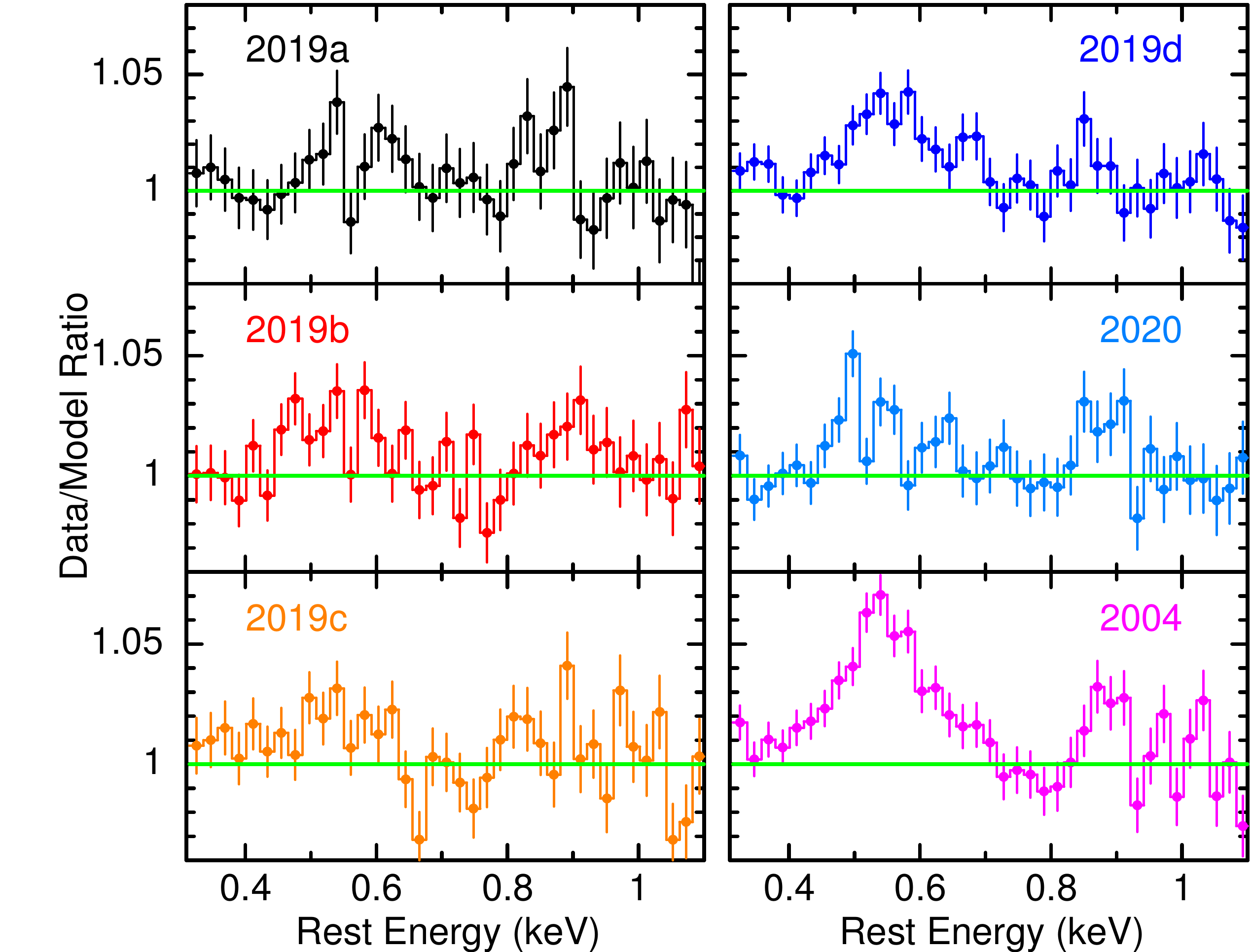}}
\rotatebox{-90}{\includegraphics[height=9cm]{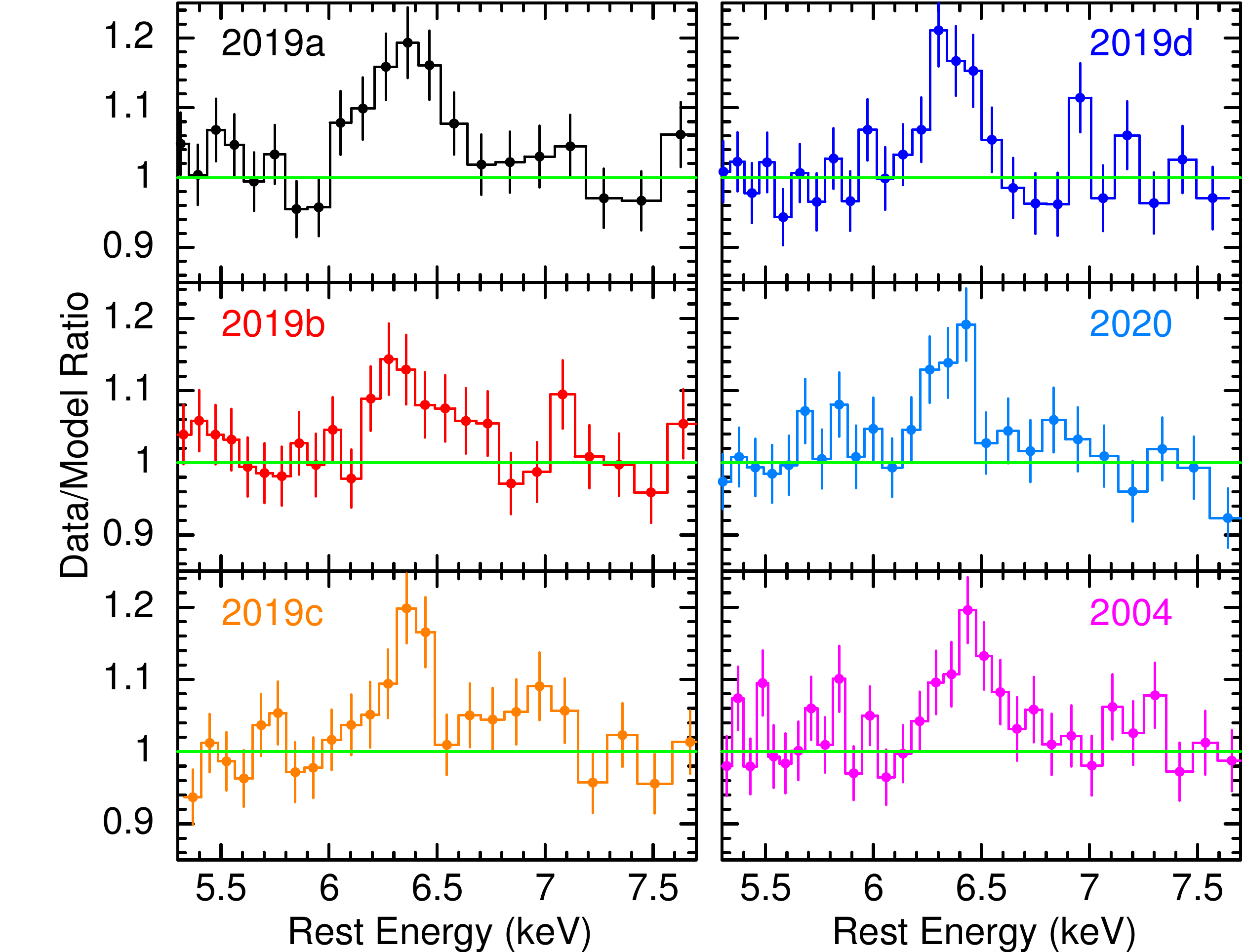}}
\end{center}
\caption{{\it Six left panels}: EPIC-pn ratio residuals of all six spectra between 0.3--1.1\,keV, compared to the two component continuum model as described in the main text. Significant excess residuals are seen 
over the 0.5--0.6\,keV O\,\textsc{vii} line region, in particular towards the 2004, 2019d, and 2020 pn spectra, where the thin filter was in place. The excess emission can be fitted with a Gaussian profile, with line fluxes as reported in Table~\ref{tab:pn}. As per the RGS analysis, the pn data show a similar trend between the line flux and the continuum, whereby the 
strongest line emission is found in the brightest (2004) observation and the weakest line flux is found in the two faintest (2019a, 2020) observations. {\it Six right panels}: Equivalent data/model ratio plots showing all six pn spectra over the iron K band, where the iron line emission is less variable.}
\label{fig:pnratio}
\end{figure*}

Against this continuum model, the six observations show an excess between 0.5--0.6\,keV, over the oxygen emission band.  The data over model residuals are shown in Figure~\ref{fig:pnratio}. 
The excess is most prominent in the brightest 2004 observation, while clear positive residuals which are, however, weaker than in 2004 are also seen in the 2019d and 2020 spectra. 
The excess is less clear in the 2019a--c spectra, where the thick filter was in place. The excess was modelled with a Gaussian emission line with a variable normalisation between the six spectra. Unlike for the higher resolution RGS spectra, it is not possible to determine changes in the profile shape between the observations and thus the line energy and width were tied between the datasets. The subsequent best-fit line energy was $E=548\pm7$\,eV (in agreement with the mean RGS line centroid energy), while the Gaussian width was $\sigma=38\pm13$\,eV. 

The results are listed in Table~\ref{tab:pn}. The line fluxes follow the same trend as that reported in Section~3.2 of the main article and thus increase with the continuum flux. 
The two lowest line fluxes were obtained for the faintest observations; for example, in the 2019a observation, an upper-limit of $N<1.2\times10^{-4}$\,photons\,cm$^{-2}$\,s$^{-1}$ was measured, while a line flux of $N=1.2\pm0.5\times10^{-4}$\,photons\,cm$^{-2}$\,s$^{-1}$ was found in the 2020 spectrum. In contrast, the bright 2004 observation has the highest line flux of 
 $N=4.3\pm0.6\times10^{-4}$\,photons\,cm$^{-2}$\,s$^{-1}$, thus a factor of $3-4$ variability is seen in the line flux. Indeed, as is shown in Figure~\ref{fig:correlation}, the line fluxes obtained from the EPIC-pn spectra agree well with those obtained from the RGS within the measurement errors. In the EPIC-pn spectra, the O\,\textsc{vii} emission is detected in five out of six observations at $>99$\% confidence, with an upper-limit for the faintest 2019a observation. The addition of the Gaussian emission component across all six spectra led to an improvement in the fit statistic from $\chi^{2}/\nu=6007/5514$ (without the O line) to $\chi^{2}/\nu=5778.1/5506$ (with the line included). 

Finally we note that an iron K line was detected in all six spectra. The data over model ratios to each of the six spectra are also shown in Figure~\ref{fig:pnratio} (right panels). 
Measured with a Gaussian profile, the best fit line energy was $E=6.35\pm0.02$\,keV, with the 
exception of the 2004 spectrum, where the line energy is slightly higher ($E=6.44\pm0.04$\,keV). 
Unlike for the O line, no evidence was found for any variability of the 
iron line flux. This ranged between $N=1.10\pm0.39\times10^{-5}$\,photons\,cm$^{-2}$\,s$^{-1}$ (lowest, 2020 observation) to $N=1.63\pm0.45\times10^{-5}$\,photons\,cm$^{-2}$\,s$^{-1}$
(highest, 2019d): That is to say they are consistent within the errors. The line equivalent widths are also consistent, as is reported in Table~\ref{tab:pn}. This may be due to the comparatively lower continuum variability in the higher energy band at about the 40\% level.  
The line width (tied between all six observations) was found to be $\sigma=75^{+35}_{-37}$\,eV, which suggests it is just marginally resolved by the EPIC-pn detector. 
There is no evidence for the width to vary between the six observations, or for any additional line components (e.g. from He or H-like iron). Higher resolution observations, for example at calorimeter resolution with \textsc{xrism} or \textsc{athena}, are required to test whether the Fe K emission line represents a face-on disk profile, as per the O\,\textsc{vii} line in the RGS spectra. 

\section{RMS variability analysis}

In order to further investigate the variability of the O\,\textsc{vii} line, rms variability spectra were extracted for the RGS over the O K-shell band. There is not sufficient variability to calculate the rms spectra for each individual observation, which are of only a 40\,ks duration (e.g. Figure~\ref{fig:rgslc}). However, there is sufficient variability between each of the six observations to calculate the rms spectrum across all the observations. To do this, each of the individual RGS spectra were binned into equal constant wavelength bins, increasing the bin size compared to the mean spectra to $\Delta\lambda=0.2$\,\AA\ per bin in order to increase the signal to noise. 
The excess variance, $\sigma^{2}_{\rm XS}$, was then calculated about all six observations in each energy (or wavelength) bin. This is defined as follows: $\sigma^{2}_{\rm XS} = S^{2} - \overline{\sigma^{2}_{\rm err}}$, where $S^{2}$ is the sample variance as a function of energy and $\overline{\sigma^{2}_{\rm err}}$ is the mean square error for each energy bin. The rms spectrum is simply the square root of the excess variance versus energy.  
Errors were calculated following the prescription given in Appendix B in \citet{Vaughan03}.

Figure~\ref{fig:rms} shows the RGS rms spectrum, calculated over the 0.45--0.7\,keV band and transposed into the rest-frame of the AGN. The continuum slope of the rms spectrum was fitted with a power-law of 
index $\Gamma=2.70\pm0.12$. This is steeper than those measured in the mean RGS spectra, where the photon index below 1\,keV varies from $\Gamma=2.12\pm0.03$ (2019a) to $\Gamma=2.45\pm0.05$ (2004). It is consistent with the spectra becoming softer with an increasing flux (Figure~\ref{fig:pnbroad}) and thus the variable continuum has a steeper photon index. The lower panel of Figure~\ref{fig:rms} shows the data and model residuals of the rms 
spectrum compared to the powerlaw. Excess residuals are observed in the O\,\textsc{vii} emission band between 0.54--0.57\,keV. This can be parameterised by a broad Gaussian function, 
with an energy of $E=562\pm3$\,eV, a width of $\sigma=6.8\pm2.0$\,eV, and an equivalent width of ${\rm EW}=5.3\pm1.4$\,eV. The fit statistic improves from $\chi_{\nu}^2=68.0/42$ to $\chi_{\nu}^2=47.3/39$ upon the addition of the Gaussian component to the variable power-law continuum.

\begin{figure}[ht]
\begin{center}
\hspace{-0.6cm}
\rotatebox{-90}{\includegraphics[height=9cm]{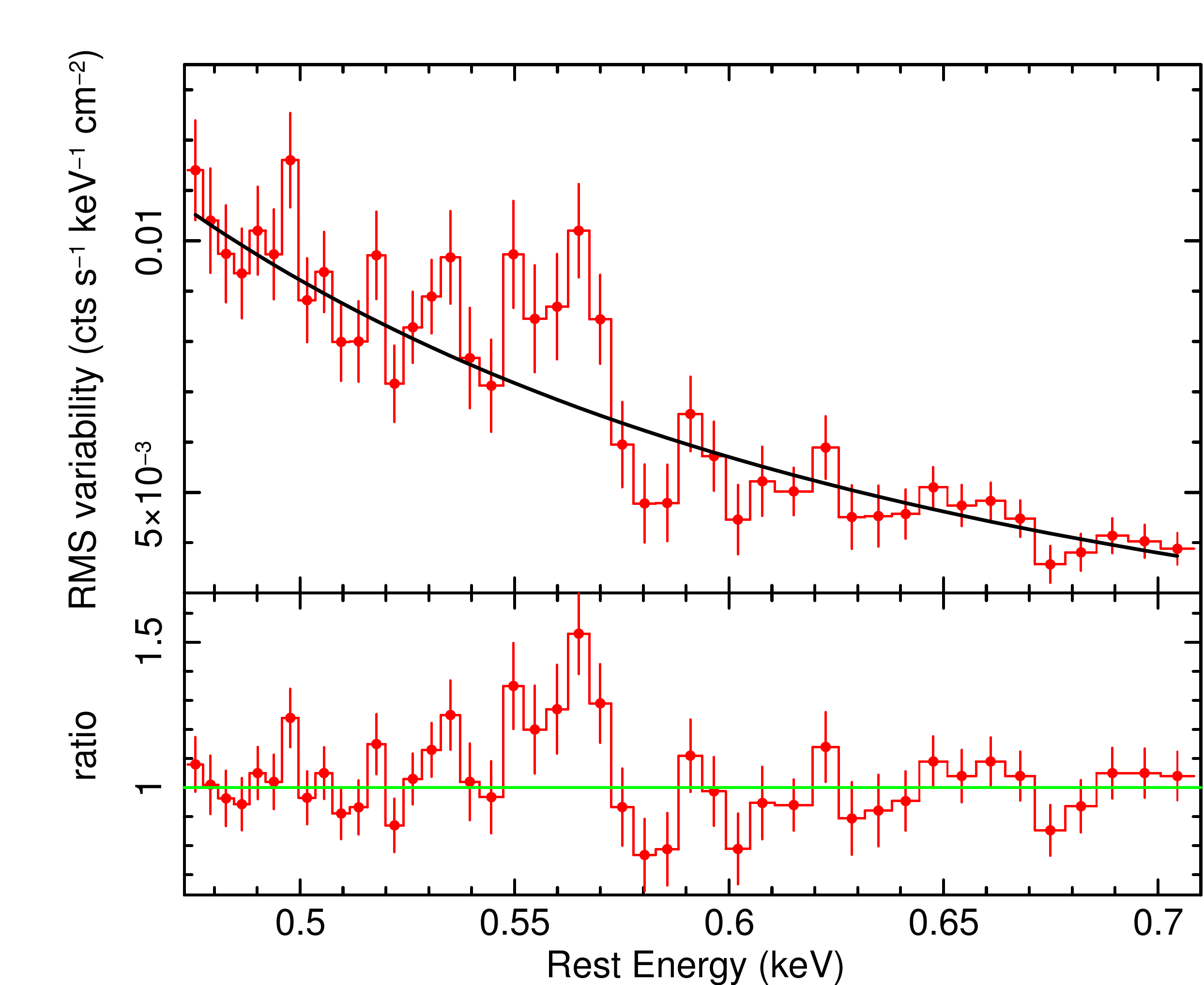}}
\end{center}
\caption{Spectrum showing the rms variability of Mrk\,110, computed across the six RGS observations in the O band. The upper panel shows the rms spectrum in counts space, compared to a power-law component (black line) for the variable 
continuum. The lower panel shows the ratio of the rms spectrum to the powerlaw model, where an excess is visible in the O\,\textsc{vii} emission band between 0.54--0.57\,keV. This confirms the line variability across all six spectra, as is presented in the main article.}
\label{fig:rms}
\end{figure} 

The variability of the O\,\textsc{vii} line versus the continuum might be predicted to follow one of three scenarios in the rms spectrum. If the line flux is less variable than the continuum (or is constant in flux), then a deficit in counts should be observed in the line band as the variability is suppressed compared to the continuum. If the line varies in proportion to the continuum, then no line residuals should be present. 
On the other hand, if the line flux is more variable than the continuum, then an excess would be predicted, which appears to be the case here. 
Regardless of the interpretation, the rms spectrum provides additional evidence of a variable component of the O\,\textsc{vii} line between observations.

\end{appendix}

\end{document}